\documentclass[aps,twocolumn,prl,tightenlines,floatfix,showpacs]{revtex4} 
\usepackage[dvips]{graphicx} \usepackage[english]{babel} 
\usepackage{amsmath} \usepackage{amssymb} \usepackage{slashed}

\def\beq{\begin{eqnarray}} \def\eeq{\end{eqnarray}} 
 \def\Re {\mbox{Re}} \def\Im {\mbox{Im}} 

 \def\c{\hspace{2pt}}

\begin{document}
\title{Conductivity in Pseudogapped Superconductors: A Sum-Rule
Consistent Preformed
Pair Scenario}
\author{Dan Wulin $^{1}$, Benjamin M. Fregoso $^{1}$, Hao Guo 
$^{3}$, Chih-Chun Chien$^{2}$ and K. Levin$^{1}$} \affiliation{$^1$James 
Franck Institute and Department of Physics, University of Chicago, 
Chicago, Illinois 60637, USA} 
\affiliation{$^3$Department of Physics, University of Hong Kong}
\affiliation{$^2$Theoretical Division, Los 
Alamos National Laboratory, MS B213, Los Alamos, NM 87545, USA} 
\date{\today} 
\begin{abstract} 
We calculate the  
dc conductivity $\sigma$ in a pseudogapped high $T_c$ 
s
perconductor within a precursor superconductivity theory which
is consistent 
with gauge invariance. Our results contain
physical effects
beyond those identified previously.
Rather than presuming that lifetime effects dominate the
$T$ dependence of transport, here
we show (consistent with growing experimental support)
that the temperature dependence of the effective carrier number
is a natural consequence of the pseudogap, and demonstrate
reasonable agreement with dc transport in the 
underdoped cuprates. 
\end{abstract}
\maketitle

Understanding the $T$ dependence of the dc resistivity, particularly near optimal doping, was 
one of the first puzzles posed by the high temperature superconductors. A number of different 
models based on spin-charge separation 
to marginal Fermi liquid phenomenology 
were invoked to explain the unusual power laws observed. With 
our increased 
understanding of the pseudogap phase (which extends over most of the phase diagram, including 
optimal doping), it has become clear that earlier theories must be 
modified to accomodate the pseudogap \cite{TimuskRMP} and related Fermi arc 
effects.
Moreover, the nature of this pseudogap is currently under heated debate
as to whether it derives from precursor and/or
fluctuation superconductivity \cite{OtherCondPapers}
or an alternative order parameter \cite{Benfatto2}.
It is therefore essential to build a systematic and properly conserving transport theory within
each of these schools and their variants.

In this paper we use a pre-formed pair, sum rule- compatible 
approach to address 
the dc conductivity $\sigma(T)$ in the cuprates. 
Our goals are to determine
(i) the effects of a pseudogap on $\sigma(T)$, (ii) 
the experimentally deduced temperature dependence
of the lifetime, (iii) 
how to understand superconducting
coherence effects in $\sigma(\omega \approx 0, T)$ \cite{Orenstein}
from above to below $T_c$,
(iv)
the role
of the Fermi arcs as well as (v) what is the source
of the ``bad metallicity" \cite{Emery} in the cuprates. 
Our approach is based on the notion that the attraction is stronger than
in BCS theory; this leads to finite center of mass pair excitations
of the normal and the condensed state.
The formal machinery used here for addressing transport is well
established \cite{Chen2,Kosztin2,OurBraggPRL}. These pair fluctuations
are a natural
consequence of the short coherence length,
as well as generally high $T_c$.
They are to be distinguished from phase fluctuations; as a mean field
approach we do not include
superconducting fluctuations which dominate in the critical region.

For illustrative purposes,
it is helpful, first, to present our expression for the dc conductivity
for $s$-wave jellium, and thereby illustrate what is the role of pre-formed or
non-condensed pairs.
We find 
\begin{eqnarray}
     \begin{array}{l}
       \sigma (T) 
     \end{array}
=\frac{1}{3 \pi^2}\displaystyle{\int_0^{\infty}}dk\frac{k^4}{m^2}
\frac{E^2_{\bf k}-\Delta_{\textrm{pg}}^2}{E^2_{\bf k}}
(-\frac{\partial f(E_{\bf k})}{\partial E_{\bf k}})
~\tau_{\sigma}(T) 
\label{eq:19}
\end{eqnarray}
Here $E_{\bf k}$ is the usual BCS dispersion,
$E_{\bf k} \equiv \sqrt{\xi_{\textbf{k}}^2 + \Delta^2_{\textbf{k}}}$
where the excitation gap consists of two contributions
from non-condensed (pg) and condensed (sc) pairs:
via $\Delta^2_{\textbf{k}} \equiv  \Delta_{pg,\textbf{k}}^2 + \Delta_{sc,\textbf{k}}^2$, 
$f(E_{\textbf{p}})$ is the Fermi function and
$\tau_{\sigma}$ is the effective lifetime.
The
gap $\Delta_{\textbf{k}}$ remains relatively T-independent, even below $T_c$, as
observed, because of the conversion of non-condensed ($\Delta_{pg,\textbf{k}}$)
to condensed
($\Delta_{sc,\textbf{k}}$) pairs as the temperature is lowered.
This transport equation is for 
the weak scattering limit and when
$\Delta_{pg}=0$ it reduces to strict BCS theory,
where below $T_c$, we consider $\omega \rightarrow 0^+$. 

Importantly, this scheme is associated
with the standard
\cite{CSTL05,FermiArcs,Levchenko}
pseudogap self energy, which we derived
even earlier within our microscopic formalism \cite{Malypapers}
\begin{eqnarray}
\Sigma_{pg}(K)=-i\gamma+\frac{\Delta_{pg,\textbf{k}}^2}{i\omega_n+\xi_{\textbf{k}}+i\gamma}
\label{eq:4a}
\end{eqnarray}
Here
$\gamma$ represents a damping 
related to
the inter-conversion of pairs and fermions.
(Rather than introducing two different $\gamma$'s in the
first and second term, we minimize the number of parameters and take them
equal.)

Physically, Eq.~(\ref{eq:19}) implies that
\textit{the 
dc conductivity is affected by
the pseudogap or pre-formed pairs, via a reduction as well as a temperature
dependence of
effective carrier number
$(n/m(T))_{\rm{eff}}$.}
The first of these arises from the Fermi function derivative, which
reduces the number of carriers through a gap effect. The second of these
appears
via the pre-factor $- \Delta_{pg}^2$ which reflects the
fact that when fermions are tied up into pairs the carrier number
is decreased. Indeed, there is an increasing experimental awareness 
\cite{AndoRes1,AndoRes2}, 
that 
``the dc resistivity of cuprates $\rho_{dc}(T)$
is governed not only by the relaxation processes but also by temperature-dependent numbers of
carriers''.

\textbf{Formalism}
A central question here is how to incorporate the widely used  Eq.(\ref{eq:4a}) into
a consistent treatment of transport.
Based on
Eq.(\ref{eq:4a}), one
can write
for the full
Green's function including condensed pair (sc) effects 
\begin{eqnarray} G_K=\Big(i\omega_n-\xi_{\textbf{k}}
+i\gamma-\frac{\Delta_{pg,\textbf{k}}^2}{i\omega_n+\xi_{\textbf{k}}+i\gamma}
-\frac{\Delta_{sc,\textbf{k}}
^2}{i\omega_n+\xi_{\textbf{k}}}\Big)^{-1}
\label{eq:3b}
\end{eqnarray}
Note that there can be no finite lifetime effects associated with the
condensed pairs.

Within a BCS-like formulation,
transport properties will involve terms
of the form $ F_{sc,K}$
which represent the usual Gor'kov functions as a product of one dressed
($G$) and one bare ($G_0$) Green's function ($GG_0$),
\begin{eqnarray}\label{eq:Fsc}
F_{sc,K} \equiv
 -\frac{\Delta_{sc,\textbf{k}}}{i\omega_n+\xi_{\textbf{k}} }
(i\omega_n - \xi_{\textbf{k}}
-\frac{\Delta^2_{\textbf{k}}}{i\omega_n+\xi_{\textbf{k}} })^{-1}
\label{eq:Gorkov}
\end{eqnarray}
To address transport we consider the EM kernel ${\bf {J}} = - \overleftrightarrow{K} \bf{A}$,
with  
$\tensor{K}(Q) = e^2\big(\tensor{n}/m\big)_{dia} + \tensor{P}(Q)$, where the paramagnetic 
contribution, given by $\tensor{P}(Q)$, is associated with the normal current from 
fermionic and bosonic excitations.  

The 
diamagnetic current 
(with electronic dispersion
$\xi_{\textbf{k}}$), 
is $(\tensor{n}/m)_{dia}=2\sum_{K}(\partial^2\xi_{\textbf{k}}/\partial \textbf{k}\partial\textbf{k})G_K$.
We integrate this expression by parts and use the 
generalized Ward identity   
to obtain \cite{Kosztin2} an alternate form (above $T_c$) 
\begin{eqnarray} 
\Big(\frac{\tensor{n}}{m}\Big)_{dia}\!\!=\!\!-2\displaystyle{\sum_K}\frac{\partial\xi_{\textbf{k}}}{\partial\textbf{k}}\frac{\partial\xi_{\textbf{k}}}{\partial\textbf{k}}\Big[G_K^2 
\!+\!\displaystyle{\sum_P}t_{pg}(P)G^2_{0,P-K}G^2_K\Big]\  
\label{eq:1} 
\end{eqnarray} 
In applying the Ward identity to arrive at Eq.~(\ref{eq:1}), 
we are considering a t-matrix approach in which the self energy can be written
as
$\Sigma_{pg}(K) 
=
\sum_Qt_{pg}(Q) G_{0}(Q-K)$, 
with
$t_{pg}(Q)=(U^{-1}+\chi_{pg}(Q))^{-1}$ where $\chi_{pg}(Q)=\sum_K
G_KG_{0,Q-K}$.
Here $U$ is the strength of an attractive
interaction which is unspecified. The presence of $GG_0$ in the non-condensed
pair t-matrix, $t_{pg}$ follows
from the $GG_0$ form of the Gor'kov function (Eq.~(\ref{eq:Gorkov})).
We define $K=(\textbf{k},i\omega_n)$
($Q=(\textbf{q},i\Omega_m)$) where $i\omega_n$ ($i\Omega_m$) is a fermionic (bosonic) Matsubara
frequency.

Eq.(\ref{eq:1})
is important because 
it has cast the diamagnetic response in the form of a two particle correlation function.
That there is no Meissner effect in the normal state is related to a precise
cancellation between the diamagnetic and paramagnetic terms. Noting
$\tensor{P}(0)=-e^2(\tensor{n}/m)_{dia}$, we may
extend $\tensor{P}(0)$ to finite $Q$ and infer that in the normal state 
\begin{eqnarray}
{\tensor{P}}(Q)&=& {2e^2}\sum_K\frac{\partial\xi_{\textbf{k}+\textbf{q}/2}}
{\partial\textbf{k}}\frac{\partial\xi_{\textbf{k}+\textbf{q}/2}}
{\partial\textbf{k}}\Big[G_KG_{K+Q}\nonumber\\
&+&\sum_Pt_{pg}(P)G_{0,P-K-Q}G_{0,P-K} 
G_{K+Q}G_K\Big].
\label{eq:7d}
\end{eqnarray}
One can alternatively \cite{Kosztin2,OurBraggPRL}
introduce the Aslamazov-Larkin and Maki-Thompson
diagrams to arrive at the above equation \cite{CSTL05}.
The condensate contribution is similarly associated with
a t-matrix.
$\Sigma(K) \equiv \Sigma_{sc}(K) + \Sigma_{pg}(K)$ and
$\Sigma(K)
=
\sum_Qt(Q) G_{0}(Q-K)$.
where $t(Q) \equiv t_{sc}(Q) + t_{pg}(Q)$. Since
$t_{sc}(Q) = - \delta(Q) \Delta_{sc,\textbf{k}}^2/T$,
this yields
the superconducting contribution in Eq.~(\ref{eq:3b}) 
\begin{eqnarray}
\Sigma_{sc}(K)
= \frac{\Delta_{sc,\textbf{k}}^2}{i\omega_n + \xi_{\textbf{k}}}
\end{eqnarray}

Then, in the same spirit as our previous \cite{Malypapers} derivation of Eq.(\ref{eq:4a}) we take note 
of the fact
that up to $T\approx T^*/2$, where $T^*$ is the pairing onset temperature 
(ie, where the pair chemical potential is small), 
$t_{pg}(Q)$ is strongly
peaked at small $P$. This leads us to identify
$\Delta_{pg}^2 = - \sum t_{pg}(Q)$ and clarifies the parameters
of Eq.~(\ref{eq:4a}). We 
rewrite Eq.~(\ref{eq:7d}), now including the usual condensate contribution as  
\begin{eqnarray}
\tensor{P}(Q)&\approx&{2e^2}\sum_K\frac{\partial\xi_{\textbf{k}+\textbf{q}/2}}{\partial\textbf{k}}\frac{\partial\xi_{\textbf{k}+\textbf{q}/2}}{\partial\textbf{k}}\Big[G_KG_{K+Q}\nonumber\\
&+&F_{sc,K}F_{sc,K+Q}
-F_{pg,K}F_{pg,K+Q}\Big]\label{eq:fullP}  
\label{eq:Fpg}
\end{eqnarray}
\begin{equation}
\rm{with}~~F_{pg,K} \equiv - \frac{\Delta_{pg,\textbf{k}}}{i\omega_n+\xi_{\textbf{k}}+ i\gamma}G_K
\end{equation}
From Eq.\eqref{eq:fullP} and
$\Re \c \sigma ^{para} (\Omega ) \c  \equiv 
\big( \frac{- \Im\c P_{xx}(\Omega)}{\Omega} \big)$,
the paramagnetic contribution to the dc conductivity is 

\begin{eqnarray}\label{eqn:gen_conductivity}
 Re \sigma^{para}(0)\!\!&\approx\!\!-\!\!\displaystyle{\lim_{\textbf{q}\rightarrow0}}\textrm{Im}\displaystyle{\sum_K}\Big[\frac{2e^2}{i\Omega_m}\Big(\frac{\partial\xi_\textbf{k}}{\partial k_x}\Big)^2\Big(G_KG_{K+Q}\nonumber\\
&\!\!-\!F_{pg,K}F_{pg,K\!+\!Q}\!+\!F_{sc,K}F_{sc,K\!+\!Q}\Big)\Big]_{i\Omega_m\rightarrow0^+}
\end{eqnarray}
In previous work in the literature 
\cite{Smith,Levchenko} only the first term involving $GG$ was included,
which was recognized  as inadequate\cite{Levchenko}.
Note also that Eq (\ref{eq:19}) follows directly 
from Eq.~(\ref{eqn:gen_conductivity}) in the limit of
small $\gamma \approx \tau_{\sigma}^{-1}$.
For consistency we rewrite Eq.~(\ref{eq:1}), also adding in the usual
BCS condensate terms 
\begin{eqnarray}
\Big(\frac{n_{xx}}{m}\Big)_{dia}\!\!&\approx&\!\!\!-2\displaystyle{\sum_K}\Big(\frac{\partial\xi_{\textbf{k}}}{\partial k_x}\Big)^2\Big[G_KG_{K+Q} 
\nonumber-F_{sc,K}F_{sc,K+Q}\nonumber\\ 
&-&F_{pg,K}F_{pg,K+Q}\Big] \label{eq:fullnm}
\end{eqnarray}
We are now in a position to demonstrate compatibility
with the important conductivity sum rule
$\int_0^{\infty}d\Omega Re \sigma(\Omega)=(1/2)e^2(n_{xx}/m)_{dia}$. Note that
$Re \sigma(\Omega)
= -{Im P_{xx}(\Omega)}/\Omega + \pi \delta(\Omega) [Re P_{xx}(\Omega) + e^2\big({n_{xx}}/m \big)_{dia}]$.
Integrating the first term over frequency we find
$= - \pi P_{xx}(0)$, while the
second (delta function) term yields a term
$+ \pi P_{xx}(0)$, which leaves only the diamagnetic
contribution and
yields the desired
sum rule.
\textit{Note that this
sum rule
is intimately connected to the
absence (above $T_c$) and the presence (below $T_c$) of
a Meissner effect}.
Importantly,
since $\big(\frac{n_{xx}}{m}\big)_{dia}$
can be
viewed as essentially independent of temperature, when
there are approximations
in evaluating the transport diagrams, it is appropriate to evaluate
the chemical potential $\mu$ based on the $T$-independence in
Eq(\ref{eq:fullnm}).

\begin{figure}
\includegraphics[width=3.0in,clip]
{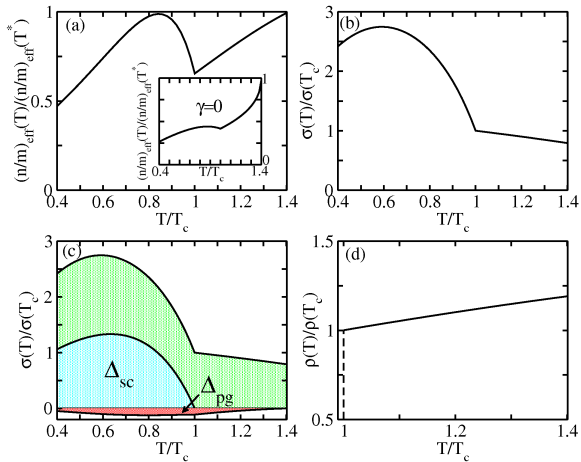}
\caption{(Color online)\label{fig1}Conductivity in an overdoped 
sample 
(a) The effective carrier number. The inset shows the effective carrier number for $\gamma/\Delta(T_c)=0$.
(b) $\sigma(T/T_c)$. 
(c) Different contributions to 
Re$\sigma(T/T_c$): the contributions due to $\Delta_{sc}$ and $\Delta_{pg}$ are labeled and the total value of $\sigma(T/T_c)$ is the unlabeled region. Panels
(b) and (d) compare favorably with experiment \cite{Orenstein,Bontemps}.}
\end{figure}

\begin{figure}
\includegraphics[width=3.0in,clip]
{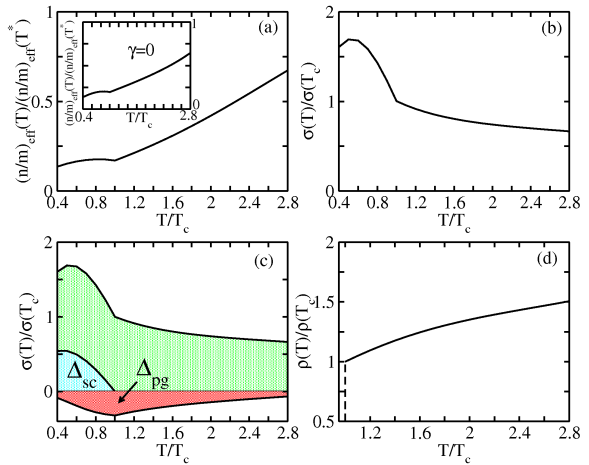}
\caption{(Color online)\label{fig2}Transport in an underdoped sample. 
(a) The effective carrier number. The inset shows the effective carrier number for $\gamma/\Delta(T_c)=0$.(b) $\sigma(T/T_c)$. 
(c) Various contributions to Re$\sigma(T/T_c)$. Corresponding (d) resistivity. Panels (b) and (d) compare favorably with experiment \cite{Orenstein,Bontemps}.} 
\end{figure}

\begin{figure}
\includegraphics[width=2.1in,clip]
{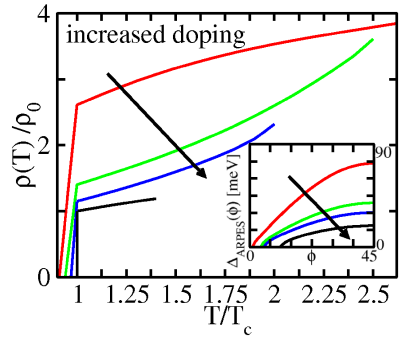}
\caption{(Color online)\label{fig3} Doping dependence of
resistivity (below $T^*$) and (inset) ARPES gaps showing Fermi arcs. The normalization $\rho_0$ is defined as $\rho(T_c)$ for the highest doping case ($\gamma/\Delta(T_c)=0.67$). The change in doping is associated with different ratios of $\gamma/\Delta(T_c)=0.07,0.20,0.35,0.67$ for increased doping.  }
\end{figure}

\textbf{Numerical Results} 
For our numerical calculations 
we have essentially two fitted parameters 
$T^\ast$ and $T_c$, representing hole doping, and these in turn constrain
the attraction $U$. We
obtain
$\mu$ and $\Delta_{\textbf{k}}(T)$ 
from the mean field equations where the mean field
transition temperature is $T^\ast$. Rather
than using a more numerically intensive approach
\cite{CSTL05}
we simplify
the decomposition of $pg$ and $sc$
contributions so that
$\Delta_{pg,\textbf{k}}(T)=(T/T_c)^{3/4}\Delta_{\textbf{k}}(T)$ for $T<T_c$ 
and 
for $ T > T_c$,  $\Delta_{pg,\textbf{k}}(T)=\Delta_{\textbf{k}}$. This parametrization of the gaps is very close 
to the full numerical solution \cite{CSTL05}.  
The values for $\gamma$ were
chosen to give rough agreement
with the observed Fermi arc lengths. 
Although it is not particularly critical, we take
the form $\gamma \propto (T/T_c)^3$ for
$T<T_c$.
This is based on fitting the temperature dependence of the
spectral function peak width
to the measured quasi-particle scattering rate \cite{Hardy1}. We define the effective carrier number
$(n/m(T))_{\rm{eff}} ~~\equiv ~~ \gamma(T) \sigma(T)$. 
It should not be surprising that the effective carrier number increases with temperature because of the
presence of an excitation gap and related
non-condensed pairs. This $T$-dependent
carrier number
is precisely what is found in a strict BCS
superconductor below $T_c$.
This contribution leads to a non-metallic
tendency with $\sigma$ \textit{increasing} with $T$ above $T_c$.

Experimentally \cite{TimuskRMP}, one finds an increasing resistivity
with $T$ and a quasi-linear $T$ dependence.
Conventionally this is explained by assuming that 
$(n/m(T))_{\rm{eff}}$ is $T$-independent, but that the 
inverse lifetime is linear in temperature. We have just shown that
due to the pseudogap, the carrier number necessarily increases with $T$. This suggests
that
$\gamma(T)$ must be a higher power than linear.
It is not implausible to assume that the gapless Fermi arcs lead to
a
more conventional,
Fermi liquid behavior; thus we choose $\gamma \propto (T/T_c)^2$.

In Fig. \ref{fig1} we plot results for a prototypical overdoped system
with $T^*/T_c = 1.4$ and 
$\gamma/\Delta(T_c)=0.67$. 
In the upper left inset we 
also plot
$(n/m(T))_{\rm{eff}}$ in the limit that $\gamma \rightarrow 0$, so that
there is no distinction between condensed and non-condensed
pairs and Fermi arcs are not present.
The lowest $T$ considered is 
$T/T_c = 0.4$ where the numerics are well controlled
and impurity effects are less important. The highest temperatures
considered correspond to $T^\ast$, above which the simple
form for the fermionic self energy
of Eq.~(\ref{eq:4a}) is no longer valid.
It
should be noted that 
both $\sigma$ and $(n/m(T))_{\rm{eff}}$
vanish at strictly zero temperature, as we do
not include a self consistent treatment of $d$-wave localization \cite{Lee1993}.
Plotted in 
Fig.\ref{fig1}(a) is the
corresponding effective carrier number
$(n/m(T))_{\rm{eff}}$. From this figure 
one can see that the carrier number begins
to rise once $T$ exceeds $T_c$.
Because the difference
between the inset and main figure are not dramatic, we see that
the arcs are not crucial for this aspect of transport.
In
Fig.\ref{fig1}(b) we plot $\textrm{Re}\sigma(T/T_c)$, which
exhibits a maximum below $T_c$, as has been 
been observed experimentally \cite{Orenstein}. 
This is associated
with
the onset of order via $\Delta_{sc}$, which increases
$\sigma$ (Eq.~(\ref{eqn:gen_conductivity})).
Above $T_c$, we see that
the conductivity of the normal state is appropriately metallic, but
suppressed by the excitation gap. In
Fig.\ref{fig1}(c) we plot and label the $\Delta_{sc}$ and $\Delta_{pg}$ components
of Re$\sigma$ arising from the three different contributions on the right
in
Eq.~(\ref{eqn:gen_conductivity}) as well as the total dc conductivity.
The ($pg$) contribution from the 
non-condensed pairs lowers the conductivity because
the presence of
non-condensed pairs means fewer fermions are available for 
\textit{dc} transport.
The resistivity $\rho$, shown
in Fig.\ref{fig1}(d) has a nearly linear 
temperature dependence, but rises faster than $T$ 
near $T^*$ because
the effective carrier number is nearing 
its normal state value.

The counterpart
results for
an underdoped sample with 
a much larger gap and $\gamma/\Delta(T_c)=0.07$ and $T^*/T_c = 4$, are shown in 
Fig.\ref{fig2}. The effective carrier number is plotted
in Fig.\ref{fig2}(a), and the inset shows the results for $\gamma =0$. 
Above $T_c$, the curvature
of
$(n/m(T))_{\rm{eff}}$
changes to convex bowing, whereas it is concave in the overdoped case.
This is to be expected as this
larger value for $\Delta(T_c)$
suppresses the carrier number.
The conductivity plotted in 
Fig.\ref{fig2}(b) is similar to its experimental counterpart and 
the difference
between this and the overdoped case is made apparent in Fig.\ref{fig2}(c).
One sees that the contribution
from
non-condensed pairs is much larger at lower doping due to the larger pseudogap.
The larger depression in
$(n/m(T))_{\rm{eff}}$
in turn leads to a concave bowing in the resistivity,
seen in Fig.\ref{fig2}(d).

Figure 3 presents a plot of normal state resistivities 
and (inset) spectral gaps 
associated with the fermionic spectral function
for dopings that interpolate between the 
underdoped and overdoped cases. The resistivities are normalized by the value $\rho_o = \rho(T_c)$ for the
case of highest doping. We may
characterize each curve, in order of increasing doping, by the ratio 
$\gamma/\Delta(T_c)=0.07,0.20,0.35,0.67$, which lead to progressively larger
arcs. The
size of the resistivities decreases as one 
increases the hole concentration and this reflects the change in
gap size and effective carrier number.
One can see from the figure that
there is a change in the nearly linear slope with increased
doping from concave to convex bowing, which may have
been seen experimentally \cite{Bontemps}.
The spectral gaps at $T_c$, displayed in the inset of
Fig.\ref{fig3} show 
the general experimental trend \cite{TimuskRMP,Levchenko} where the
arc length increases with doping. 

\textbf{Conclusions} 
Our main observation of $T$-dependence in the carrier
number (and its implications for
dc transport) seems quite natural and general, although
the community 
focus has been on the $T$ dependence of the scattering time.
The present pre-formed pair approach is importantly demonstrably
consistent with conservation laws and 
extendable below $T_c$ (merging into a BCS ground state). 
By contrast, other papers in the
literature \cite{Levchenko,Smith,OtherCondPapers} are restricted to
$ T > T_c$ and, unlike Ref.~\onlinecite{Benfatto2} do not establish sum rule consistency.
The collapse of the Fermi arcs and the so-called two-gap physics
(through Eq.~(\ref{eq:3b}))
are also evident \cite{ChienArpes} here, as is
the anomalously low conductivity, or ``bad-metal" behavior \cite{Emery}.
The (gapless) Fermi arcs appear to be
more important for the relaxation
time than for the carrier number.

This work is supported by NSF-MRSEC Grant
0820054. 
C.C.C. acknowledges the support of the U.S.
Department of Energy via the LANL/LDRD Program.

\bibliographystyle{apsrev}
\bibliography{Review2}

\end{document}